# Substrate Effects on the Speed Limiting Factor of WSe$_2$ Photodetectors

*Christine Schedel[1], Fabian Strauß[1,2], Pia Kohlschreiber[1], Olympia Geladari[1], Alfred J. Meixner[1,2], Marcus Scheele[1,2,\*]*

[1]Institute for Physical and Theoretical Chemistry, University of Tübingen, 72076 Tübingen, Germany.

[2]Center for Light-Matter Interaction, Sensors and Analytics LISA+, University of Tübingen, 72076 Tübingen, Germany.

ABSTRACT  We investigate the time-resolved photoelectric response of WSe2 crystals on common glass and flexible polyimide substrates to determine the effect of the dielectric environment on the speed of the photodetectors. We show that varying the substrate material can alter the speed-limiting mechanism: while the detectors on polyimide are *RC* limited, those on glass are limited by slower excitonic diffusion processes. We attribute this to a shortening of the depletion layer at the metal electrode/WSe2 interface caused by the higher dielectric screening of glass compared to polyimide. The photodetectors on glass show a tunable bandwidth which can be increased to 2.6 MHz with increasing the electric field.

**INTRODUCTION**

Transition metal dichalcogenides (TMDCs) are attracting attention for the development of fast photodetectors due to high charge carrier mobilities, high on/off ratios in field-effect transistors,[1] and large layer- and dielectric-dependent exciton binding energies.[2-5] Their high mechanical stability makes them ideal candidates for flexible, transparent optoelectronic applications.[6] The transport properties of TMDCs are tunable by extrinsic factors, such as the flake thickness and electrode material. Specifically, for semiconducting $WSe_2$ it has been shown that the carrier type can be tailored from p-type via ambipolar to n-type with increasing film thickness.[7-9] The speed of TMDC photodetectors is tunable via the electrode material[10-13], surface oxidation/doping[14-16], strain[17] and the applied gate voltage[13, 16, 18], typically evidenced by the rise and fall time of the photoresponse towards square pulse illumination. An underexplored factor in this regard is the dielectric environment provided by the substrate. In view of the small thickness of TMDC devices, the effect of the underlying substrate is anticipated to be strong, which has indeed been demonstrated for related properties, such as the excitonic behavior of TMDCs, their photoluminescence[19-22] as well as the carrier mobility[11, 23]. For $MoS_2$ monolayers, different surface treatments and growth techniques on a $Si/SiO_2$ surface resulted in fall times in the range of 0.3 – 4000 s, which was attributed to the varying hydrophobicities of the substrates.[13] We are unaware of comparable studies for $WSe_2$, in particular beyond $Si/SiO_2$ as the substrate, which is problematic for investigating the photodetector speed due to a convolution with the fast photoresponse of the underlying silicon.[24]

Here, we investigate the photoresponse of $WSe_2$ photodetectors on common glass (large dielectric screening) as well as flexible polyimide substrates (weak dielectric screening) to evaluate the effect of these substrates on the speed of the devices. To this end, we study both, the steady state photoresponse towards 635 nm square pulse illumination and the non-steady state photoresponse towards 636 nm and 779 nm impulse illumination. We show that the choice of substrate material can entirely change the speed-limiting mechanism of the devices, e.g. whether the photodetectors are resistance-capacitance- (*RC*-) or diffusion-limited. We ascribe this to a changed depletion width at the metal/WSe$_2$ interface due to the varied dielectric surrounding. For WSe$_2$ detectors on glass (with large dielectric constant), we find a tunable 3 dB bandwidth, which can be accelerated up to about 2.6 MHz by increasing the applied electric field.

**RESULTS**

The investigated WSe$_2$ flakes, from now on referred to as "crystals", had a typical thickness of 30 – 50 nm, corresponding to at least 40 layers[4], cf. Figures 1a and SI1. On polyimide, this thickness resulted in a weak photocurrent, which is why we had to apply thicker crystals of 100 – 200 nm for the rise- and fall-time measurements to achieve a sufficient signal-to-noise ratio. However, with 40 layers and more for all crystals, we assume bulk behavior for all detectors and neglect the differences in height when investigating the speed of the response.

The dark currents of all photodetectors under atmospheric conditions are shown in Figure 1b, where the devices on glass are depicted in black and the ones on polyimide in red. The significantly reduced dark current ($I_{dark}$) of the polyimide detectors, regardless of thickness, is promising to achieve a high on/off ratio $\frac{on}{off} = \frac{I_{Photo}}{I_{dark}}$ of the devices, with photocurrent ($I_{Photo}$).

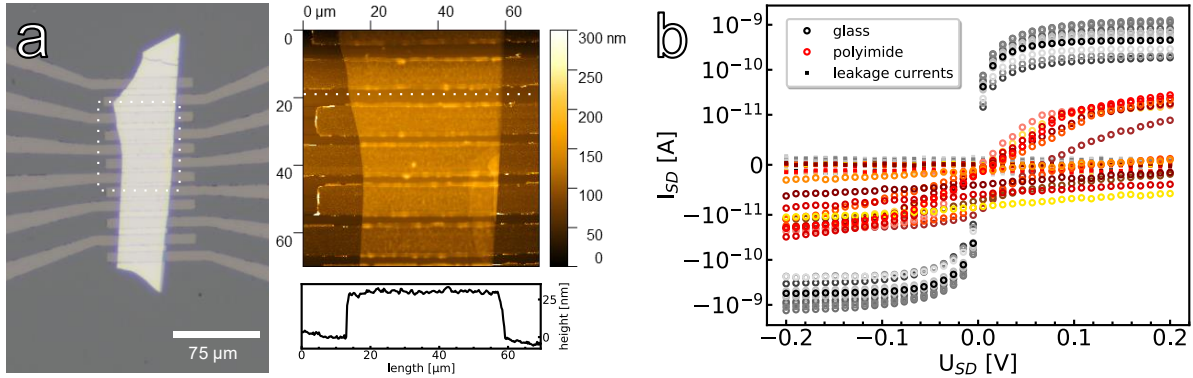

**Figure 1**. a) Light microscope and AFM image of a typical WSe$_2$ photodetector on glass with 2.5 µm / 5 µm × 80 µm bottom gold electrodes. The white box indicates the AFM cutout, the white line indicates the position of the line cut for the height profile. b) Dark currents of all WSe$_2$ photodetectors on glass (black) and polyimide (red) under atmosphere.

**Steady state (or "square pulse") photoresponse on glass and polyimide substrates.**

Square pulse illumination, reaching steady state photocurrent, examines whether geometry-related speed limitations, e.g. due to the *RC* time constant, limit the speed of the detector. The photocurrent signals were measured under various source drain biases from – 200 mV to 200 mV under atmospheric conditions and showed a nonlinear dependence of the applied bias, cf. Figure 2 (insets) and Figure S2, similar to previously reported WSe$_2$ photodetectors.[11,25]

The photoresponse of a WSe$_2$ crystal on polyimide and one on glass towards a 635 nm 10 kHz square pulse illumination (≤ 12 mW) for different applied electric fields is shown in Figures 2a and 2b, respectively. There is a significant difference in the response of glass detectors compared to polyimide devices, while the absolute photocurrents of the devices are in comparable ranges. On polyimide, no field dependence on both rise and fall times (90 % – 10 %) can be observed which is a clear indication of *RC* limitation of the devices.[24] The rise / fall times are in the range of 500 ns up to a few microseconds regardless of the applied bias and channel length, cf. Figure S3. On glass, on the contrary, the rise and fall times decrease with increasing applied voltage. They can be accelerated from approx. 6 µs when ± 50 mV is applied down to 1 µs for ± 200 mV bias applied, cf. Figures 2c and 2d. This places the WSe$_2$ photodetectors in this work among the fastest reported TMDC photodetectors.[11, 25-31] No channel length dependence is measurable. Furthermore, the rise / fall times of the glass detectors are independent of the actual photocurrents, in other words, of the photoresistances $R_{illum}$. Figure 3 shows a typical example for the photoresponse of one WSe$_2$ detector on glass with different photocurrent heights induced by a 635 nm laser at 10 kHz repetition rate. Both, the existing field dependence and the lack of photoresistance dependence of the rise / fall times, are a clear indication that the glass detectors are not limited by the $R_{illum}C$ constant of the device.

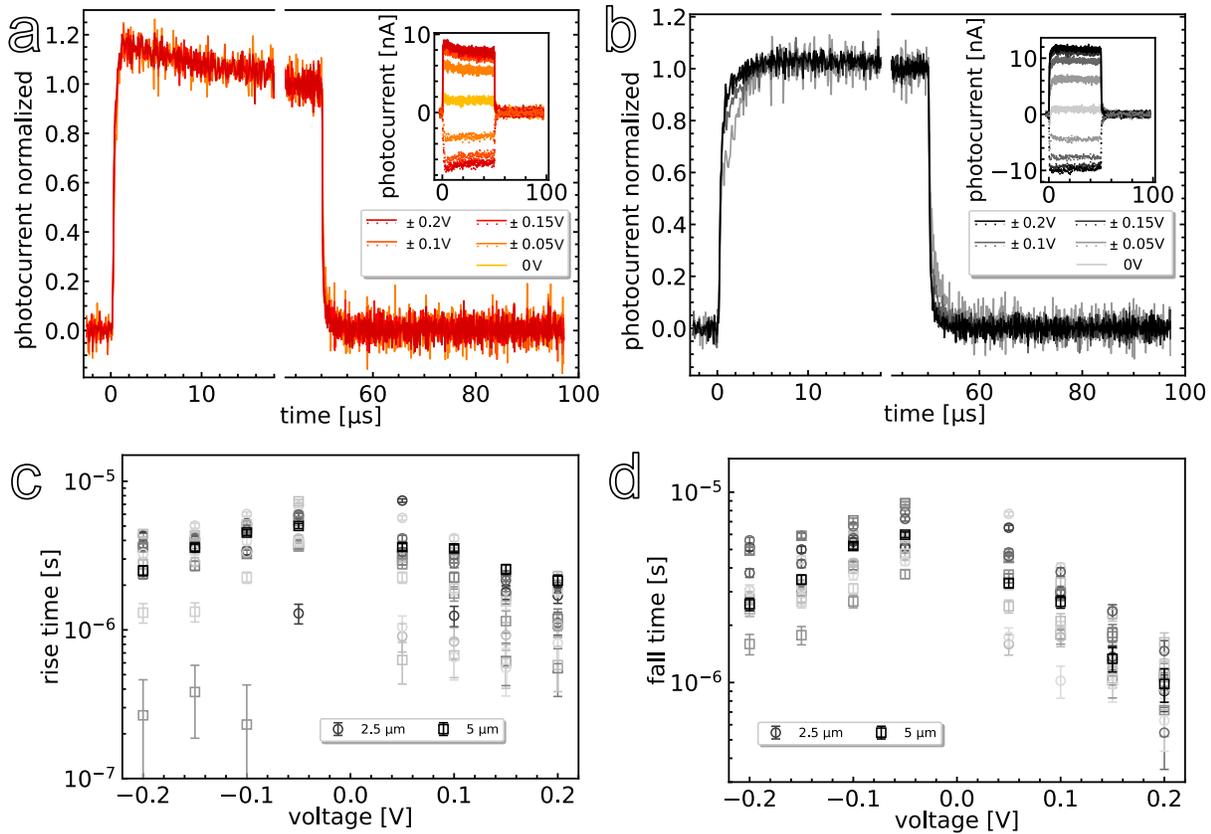

**Figure 2.** a) Typical normalized photoresponse of a WSe$_2$ crystal on a) polyimide (channel length: 2.5 µm) and b) glass (channel length: 5 µm) towards a 635 nm 10 kHz square pulse laser at different voltages. Inset: absolute photoresponses. Correlation between measured c) rise times, d) fall times vs. applied voltage for all glass devices (one color = one sample).

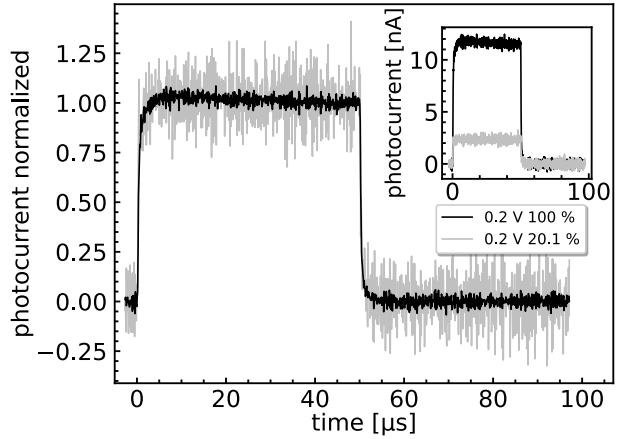

**Figure 3.** Normalized photoresponse of a WSe$_2$ detector on glass towards a 635 nm 10 kHz square pulse laser with different laser intensities. Inset: absolute photoresponse. Exemplarily shown on a 5 µm device.

To identify the $R_{illum}C$ times, we calculated the expected capacitances ($C$) of the devices according to $C = L(N-1)\varepsilon_0(1+\varepsilon_r)\frac{K(k)}{K(k')}$,[32] with the finger length ($L = 80$ µm), the number of electrode fingers ($N = 2$), the vacuum permittivity ($\varepsilon_0$), the dielectric constant of WSe$_2$ material ($\varepsilon_r$ taken as 20 for the in-plane dielectric constant at 1.95 eV incident photon energy)[33, 34], the complete elliptical integral of first kind ($K$) with $k' = \sqrt{1-k^2}$ and $k = \cos\left(\frac{\pi}{2}\left(1-\frac{w}{w+g}\right)\right)$, and the width of the electrode fingers ($w = 10$ µm). For channel lengths ($g$) of 2.5 µm and 5 µm, we therefore expect the capacitances of the devices to be about 24 fF and 19 fF, respectively. With these theoretical capacitances, we calculated the $R_{illum}C$ times. In a comparison with the measured rise times, see Figure 4, it is evident that the $R_{illum}C$ times are expected to be faster than the actual rise times on glass, whereas on polyimide, they are identical.

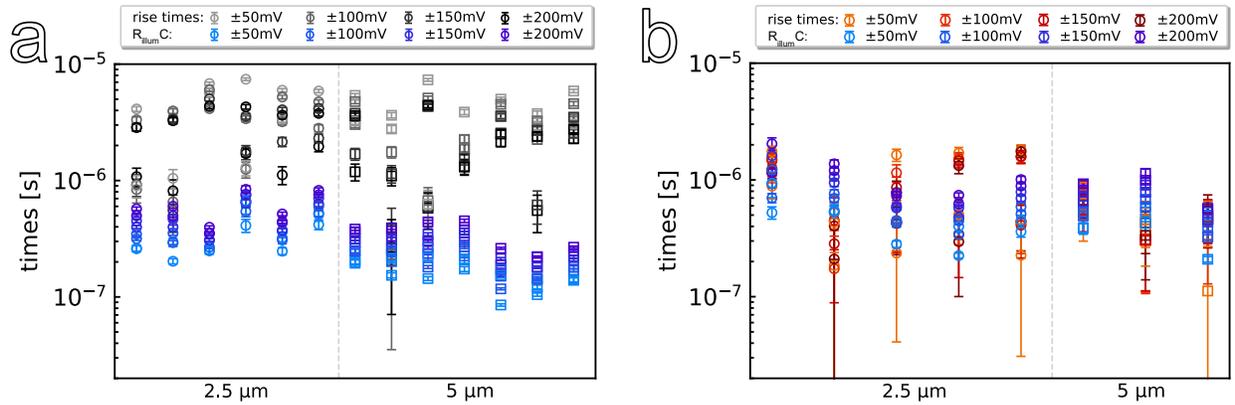

**Figure 4.** Calculated $R_{illum}C$ time and measured rise times on a) glass and b) polyimide, x-axis: electrode gap (μm × 80 μm), contacts listed.

The on/off ratio of WSe$_2$ photodetectors on glass is $18 \pm 10$ (median and standard deviation, with the highest on/off ratio of 44), whereas it is significantly higher on polyimide with approx. 325. The on/off ratios on polyimide show large standard deviations and can be as high as 3603, but also a value of only 24 has been measured once.

**Non-steady state (or "impulse") photoresponse on glass and polyimide substrates.**

Impulse illumination is used for optical communication, i.e., the response of a photodetector to a delta-shaped laser pulse is critical and determines the bandwidth of the device. The normalized impulse photoresponse of a typical WSe$_2$ photodetector on glass towards different applied biases under atmospheric conditions is shown in Figure 5a. With higher applied electric field, the fall time of the photoresponse is strongly reduced, which results in a tunable 3 dB bandwidth of the device, cf. Figure 5b. The values typically range from $132 \pm 53$ kHz under 50 mV applied up to $2.57 \pm 1.55$ MHz when 200 mV are applied.

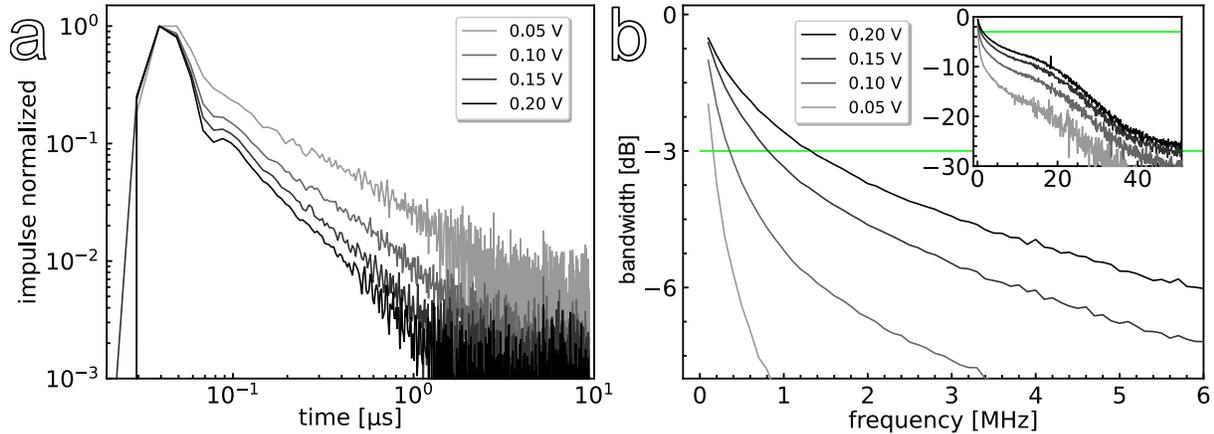

**Figure 5.** a) Normalized impulse response of a typical WSe$_2$ glass device (5 µm channel), under 636 nm 100 kHz illumination and different electric fields applied. b) Corresponding bandwidth of these impulse spectra, showing the field effect towards the impulse response in the frequency range: The 3 dB bandwidth increases from 160 kHz up to 1.32 MHz, for 0.05 V up to 0.20 V applied, respectively. Inset: full range of the bandwidth spectrum.

On the other hand, no field dependence is visible for polyimide-based photodetectors, see Figure 6a. As indicated by the reduced signal to noise ratio, those detectors exhibit a reduced photocurrent signal compared to the glass ones, also shown in Figure S4. Some of the polyimide photodetectors do not display a measurable impulse photoresponse at all, which contrasts with their high on/off ratio under steady state illumination. We calculate the 3 dB bandwidth to about $619 \pm 365$ kHz (the high standard deviation is due to the reduced signal to noise ratio).

The impulse responses are identical under 636 nm and 779 nm illumination under otherwise identical conditions for all detectors, as exemplarily shown in Figure 6b for a glass device.

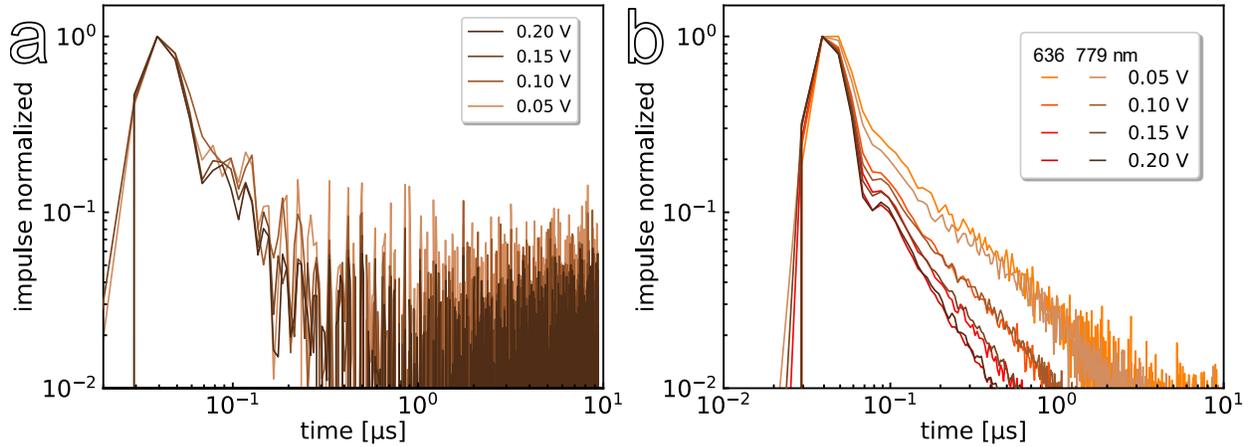

**Figure 6.** a) Normalized impulse response of a typical WSe$_2$ polyimide device (2.5 μm channel), under 779 nm 100 kHz illumination and different electric fields applied. b) Normalized impulse responses of a typical WSe$_2$ glass device (5 μm channel), under both 636 nm and 779 nm 100 kHz illumination.

**Atmospheric influences**

Transition metal dichalcogenides can be doped with a variety of substances, including physisorbed oxygen, which alters the conductivity of the material.[35, 36] Thus, not only the bottom substrate but also the top adjacent substances have to be taken into consideration. We found that photodetectors prepared under atmospheric conditions but measured under nitrogen exhibit a strongly increasing conductivity over several hours in the dark when the WSe$_2$ crystals are attached to glass substrates, see Figure S5. In contrast, the conductivity remains nearly constant when the flakes are transferred to polyimide substrates. Therefore, to exclusively study the influence of the substrates on the photodetectors and to ensure constant dark resistances, we decided to perform our investigations under ambient conditions. Moreover, chemisorbed oxygen saturates chalcogen vacancies commonly present on TMDCs, removing electron trap states and carrier scattering centers. This increases the electron mobility and photoresponse.[37] In addition, it has already been shown that by oxygen doping of ambipolar WSe$_2$ (surface oxidation) the square pulse photoresponse can be accelerated considerably.[14] We therefore concentrated on keeping the atmospheric conditions as identical as possible.

**DISCUSSION**

For a complete investigation of the photodetector performance regarding the speed of the device, both steady state studies, using square pulse illumination, and non-steady state examinations, using impulse illumination, have to be considered.

**Steady state.** Steady state data can be used to determine whether capacitive (dis-)charging of a photodetector limits its speed of response, which is defined by the *RC* time. This becomes evident by a lack of field strength dependence and channel length dependence of the rise and fall times (10 % – 90 %) towards square pulse illumination.[24] WSe$_2$ photodetectors on polyimide exhibit such an *RC* limitation. It can be seen by the constant rise / fall times for different applied voltages, cf. Figures 2a and SI3, and is confirmed by the consistency of the calculated *RC* time with the measured rise time, see Figure 4b.

On glass, on the other hand, we observe a field-dependence of the rise and fall times, see Figure 2b, and there is no dependence on the photoresistance, cf. Figure 3. This clearly argues against a *RC* limitation, and the measured rise time is indeed slower than the calculated *RC* time.

Another parameter that can limit the performance of a fast photodetector is the transit time $\tau_{drift} = \frac{d^2}{\mu V}$, with the channel length (*d*) and the charge carrier mobility ($\mu$). However, on glass the photodetectors do not exhibit a measurable dependence of the rise / fall time on the examined 2.5 µm and 5 µm channel length devices, cf. Figures 2b, 2c and 4a. This speaks against a transit time-limitation of the detectors.

Therefore, another process must be causing the time limitation of the glass devices. The Schottky barrier is a very important parameter for TMDC photodetector performance[11, 12, 25, 28, 38, 39] and photocurrents can primarily be detected in the depletion region at the interface between the crystals and the electrode.[11, 12] The depletion region is responsible for exciton dissociation, can extend over several micrometers and can be tuned by the applied gate voltage.[12] The Schottky barrier of trilayer $WSe_2$ on a $Si/SiO_2$ device, for example, can be varied from 60 meV to almost 0 eV by the applied gate voltage.[25] A gate controlled modulation of the metal-semiconductor barrier also explains the transition from a Schottky to an ohmic contact in multilayer $WSe_2$ FET.[39]

We suggest that the higher dielectric constant of glass with 4.5-8[40] compared to polyimide with 3.7[41] induces a shortening of the depletion width by stronger polarization and dielectric screening effects. If the depletion width in the crystals on glass is shortened below the channel width of 2.5 µm, slow excitonic diffusion begins to play a role. $WSe_2$ has a high spin-orbit coupling effect which leads to a complex excitonic behavior, and photoluminescence experiments indicate that charged excitons, so-called trions, appear in $WSe_2$.[19, 20, 42] Since trions are charged, their diffusion can get accelerated by an externally applied field. We do not observe a channel length dependence for this excitonic diffusion, as it is limited to only short regions. The exciton diffusion length in $WSe_2$ at room temperature is as short as 380 nm[43] (several 10 nm thick flakes, comparable to the samples investigated here) and the trion diffusion length is expected to be even shorter.[44, 45] Thus, its influence should be similar for all devices shown here. However, detailed spatial investigations of the photocurrent are beyond the scope of this work.

Substrates can influence the responsivity of photodetectors.[46, 47] This is apparent in the on/off ratio of the polyimide detectors being about one order of magnitude higher compared to the glass detectors and can be explained by the strongly reduced dark current of $WSe_2$ on top of polyimide, cf. Figure 1b. Inefficient travelling of carriers due to enhanced carrier scattering occurs in devices on rough surfaces (surface roughness: $30.7 \pm 11.6$ nm on polyimide, vs $5.5 \pm 0.7$ nm on glass, from AFM measurements, not shown here).[5, 48, 49] To prevent this, atomically flat hexagonal boron nitride is often inserted between the substrate and the active material in 2D optoelectronic devices which improves the charge carrier mobility significantly.[11, 48]

**Non-steady state.** The electrical 3 dB bandwidth is tunable for glass devices with the applied electric field and reaches values as high as 2.6 MHz under 200 mV bias. We attribute this to the acceleration of slowly diffusing trions in a not fully depleted channel, comparable to the situation under steady state illumination. On the contrary, the bandwidth on polyimide devices is only around 620 kHz and cannot be tuned, which can be ascribed to the $RC$ limitation of the device.

The photoresponse towards impulse illumination is strongly reduced on polyimide substrates compared to glass substrates, cf. Figure S4a. The substrate material affects the excitonic behavior due to the dielectric screening effect, and the exciton binding energy can vary.[19] As the binding energies of both excitons and trions in $MoS_2$ decrease with increasing dielectric constant,[50] a similar effect is plausible for $WSe_2$, which would result in a lower excitonic binding energy on glass (dielectric constant of 4.5-8[40]) compared to polyimide (3.7[41]). It could result in a reduced exciton separation and photocurrent generation efficiency on polyimide during the very short time scale of the laser pulse. Moreover, charge carriers can travel more efficiently on smooth substrates with minimum dangling bonds, minimum impurities, thus minimum charge taps, and minimized surface roughness and due to an enhanced dielectric screening.[5, 16, 48, 49] The increased scattering on the rough surface of polyimide substrates (30.7 $\pm$ 11.6 nm, vs 5.5 $\pm$ 0.7 nm on glass) and its smaller dielectric constant thus limit electric transport further. The thinner the $WSe_2$, the stronger are all of these effects. The crystals here are relatively thick with more than 30 nm, but due to the bottom contact geometry, we assume that substrate roughness and dielectric screening still play a major role. We expect the highest fields directly at the substrate-$WSe_2$ interface, thus, the lower

WSe$_2$ layers directly in touch with the substrate are expected to influence the photoresponse strongly.

The difference between the small impulse photocurrent compared to the high on/off ratio under steady state conditions on polyimide can be explained by the inclusion of the dark current. It is considered in the on/off ratio but not in the time-resolved measurements, in which only the photocurrent is registered. The reduced impulse photoresponse of polyimide detectors is problematic as it limits their applicability for photodetection in fast optical switches for data transmission.

**Atmospheric influences**. The investigated WSe$_2$ samples are at least 30 nm thick, which should lead to n-type conductance based on previous reports.[7, 8] Under atmosphere, contact with oxygen and water deprives the crystals of electrons and greatly decreases the electron mobility,[35] which lowers the conductivity compared to studies in a nitrogen environment. If detectors fabricated under atmospheric conditions are placed in a nitrogen atmosphere, oxygen probably desorbs continuously over several hours, resulting in an increasing conductivity for the glass photodetectors, cf. Figure S5. This effect is hardly observed for the polyimide devices which can possibly be attributed to the significantly higher surface roughness of polyimide film (30.7 $\pm$ 11.6 nm) compared to glass slides (5.5 $\pm$ 0.7 nm). On polyimide, more air inclusions can be trapped between the substrate and the WSe$_2$ crystal, such that exchange with nitrogen does not occur and the conductivity remains constant.

**CONCLUSION**

We have examined the time-resolved photocurrent response of WSe$_2$ crystals on polyimide (small dielectric constant) and glass substrates (large dielectric constant). We show that by changing the substrate material, the speed-limiting mechanism of the device can be changed, with a *RC* limitation for crystals on polyimide and a diffusion-limitation for those on glass. We ascribe this to a shorter depletion width at the metal/WSe$_2$ interface on glass due to the higher dielectric screening. The diffusion limitation of photodetectors on glass leads to a tunable bandwidth, which can be increased to about 2.6 MHz with increasing electric field.

**EXPERIMENTAL METHODS**

**Device Preparation**. The identical electrode geometry (2.5 / 5 μm × 80 μm, cf. Figure 1a) was photolithographically prepared on 0.125 mm polyimide foil (DuPont$^{TM}$ Kapton® HN) and glass slides (Duran Wheaton Kimble; soda-lime glass). The glass slides were cleaned in an ultrasonic bath in a potassium hydroxide solution with 30 % - hydrogen peroxide solution added after five minutes. The cleaned glass substrates were coated with hexamethyldisilazane. The polyimide foil was cleaned in the ultrasonic bath with acetone and isopropanol. Both substrate types were spin-coated with maN-405 (micro resist technology) photoresist (3000 rpm; 30 s). The photoresist was then soft-baked on a hotplate (100 °C) for one minute. Optical lithography was performed on a μMLA maskless aligner from Heidelberg Instruments (dose: 1000 mJ/cm$^2$). After development for three minutes in maD-331/S (micro resist technology), the substrates were metallized with 2 nm titanium and 8 nm of gold in a PLS570 evaporator. Finally, lift-off was done in acetone.

WSe$_2$ flakes were mechanically exfoliated from a WSe$_2$ crystal (2Dsemiconductors Inc. USA) using scotch tape (Scotch Magic$^{TM}$ Tape) and transferred via a polydimethylsiloxane stamp (PF Gel Film®, Teltec GmbH) to the prepared glass and polyimide substrates, see Figure S6. The procedure was carried out under atmosphere. The thickness of the exfoliated flakes was investigated via atomic force microscopy with a Bruker MultiMode 8-HR in contact mode, cf. Figure S1.

**Electrical Measurements**. Devices were investigated using a Keithley 2634B System Source Meter in a Lake Shore Cryotronics probe station at room temperature and under ambient conditions and/or $N_2$. The device electrodes were contacted with tungsten two-point probes.

**Transient Photocurrent**. The photocurrent performance of the devices was investigated in the probe station at room temperature and under ambient conditions. The device electrodes were contacted with 50 Ω matched tungsten two-point probes and 40 GHz coaxial cables as short as possible. A FEMTO DHPCA-100 current amplifier preamplified the current, and the Periodic Waveform Analyzer of a Zurich Instruments UHFLI lock-in amplifier was used to measure the photocurrent by averaging over 2G samples. The signals were background corrected. The lock-in amplifier limits the time resolution of the setup to 600 MHz.

The steady state photocurrent was examined under square pulse illumination. A fast switchable laser driver (FSL500, PicoQuant) with a laser rise time of < 0.5 ns together with a 635 nm laser diode with an output power of ≤ 12 mW was externally triggered at 10 kHz with a Hewlett Packard 33120A arbitrary waveform generator. Transient photocurrent measurements were carried out using impulse illumination with a picosecond pulsed laser driver (Taiko PDL M1, PicoQuant) together with laser heads for 636 nm and 779 nm illumination with a FWHM of the pulses of < 500 ps. A 100 kHz repetition rate was selected with average output powers of 22 µW. The bandwidths of the photodetectors are determined via the power spectrum $P(\omega)$, by fast Fourier transformation FFT of the impulse response f(t): $P(\omega) = |\text{FFT}(f(t))|^2$.[51] The specified laser powers are attenuated due to inefficient coupling into the optical fiber, scattering, decollimation of the beam, etc. The laser spot was not focused and illuminated an area larger than the actual active region of the photodetector.

ASSOCIATED CONTENT

**Supporting Information**. Figure S1: Characterization of the investigated WSe$_2$ flakes, Figure S2: Voltage dependent steady state photocurrent, Figure S3: Voltage dependent rise / fall time of WSe$_2$ on polyimide, Figure S4: Absolute impulse photoresponses, Figure S5: Dark currents under atmosphere and nitrogen, Figure S6: Mechanical exfoliation and stamping process.

AUTHOR INFORMATION

**Corresponding Author**

*To whom correspondence should be addressed: marcus.scheele@uni-tuebingen.de

**Author Contributions**

The manuscript was written through contributions of all authors. All authors have given approval to the final version of the manuscript.

**Funding Sources**

Financial support of this work has been provided by the European Research Council (ERC) under the European Union's Horizon 2020 research and innovation program (grant agreement No 802822) as well as by the Deutsche Forschungsgemeinschaft (DFG) under grant SCHE1905/9-1.

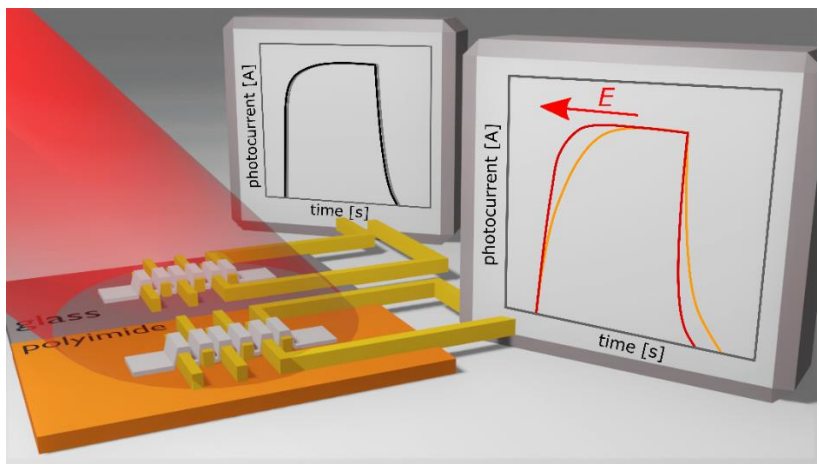

Table of content graphic.

# Supporting Information

# Substrate Effects on the Speed Limiting Factor of WSe₂ Photodetectors

*Christine Schedel[1], Fabian Strauß[1,2], Pia Kohlschreiber[1], Olympia Geladari[1], Alfred J. Meixner[1,2], Marcus Scheele[1,2,*]*

[1]Institute for Physical and Theoretical Chemistry, University of Tübingen, 72076 Tübingen, Germany.

[2]Center for Light-Matter Interaction, Sensors and Analytics LISA+, University of Tübingen, 72076 Tübingen, Germany.

Corresponding author: marcus.scheele@uni-tuebingen.de

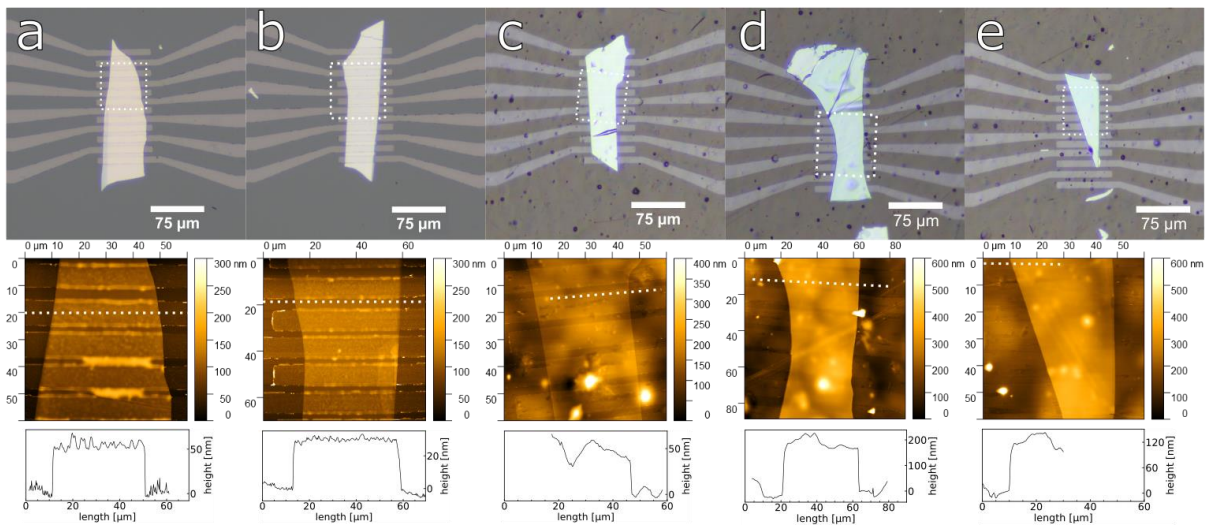

**Figure S1.** Light microscope and AFM images of all investigated WSe$_2$ crystals on a, b) glass and c-e) polyimide. The white boxes indicate the AFM cutouts, the white lines show the position of the line cuts for the height profiles.

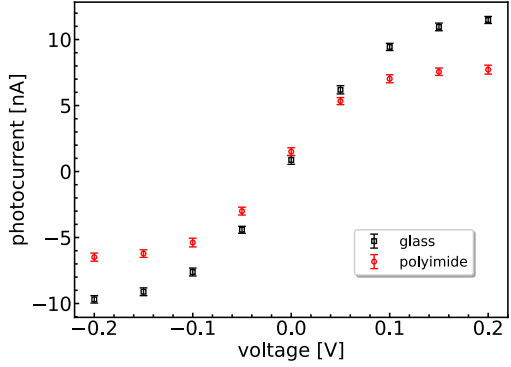

**Figure S2.** Steady state photocurrent vs source drain bias of one WSe$_2$ detector on glass (black, 5 µm device) and one on polyimide (red, 2.5 µm device) upon 635 nm illumination ($\leq$ 12 mW).

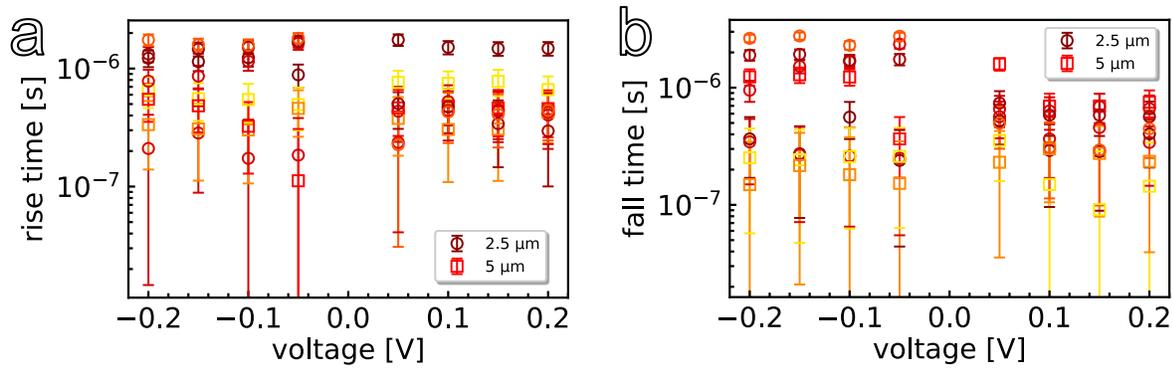

**Figure S3.** Correlation between measured a) rise times, b) fall times vs. applied voltage for all polyimide devices (one color = one sample).

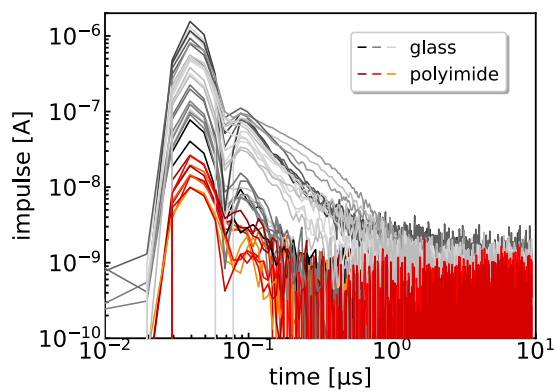

**Figure S4.** Absolute impulse responses of WSe$_2$ flakes on glass (grey) and polyimide (red) with 0.20 V applied, towards 779 nm 100 kHz illumination.

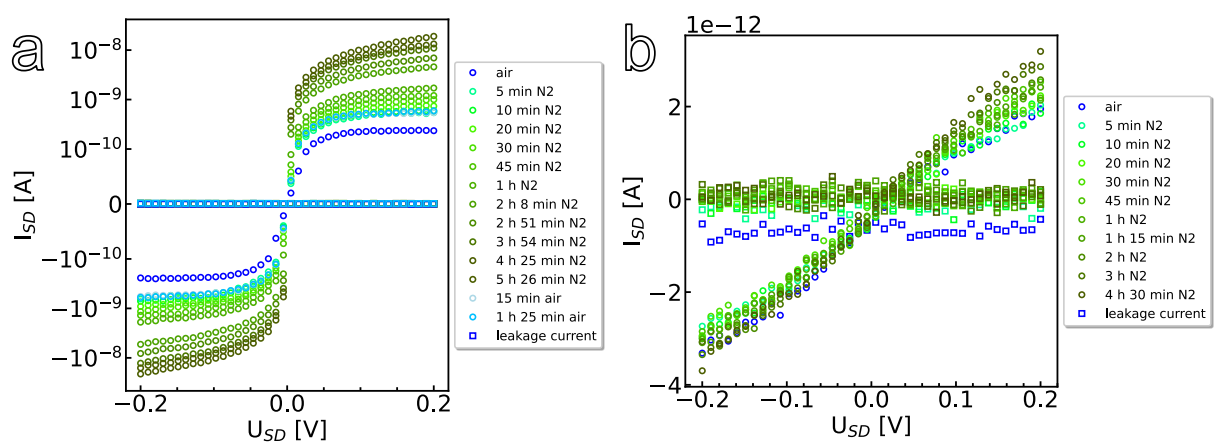

**Figure S5.** Dark current evaluation under atmosphere and some time in nitrogen, for typical 5 μm WSe$_2$ devices on a) glass and b) polyimide.

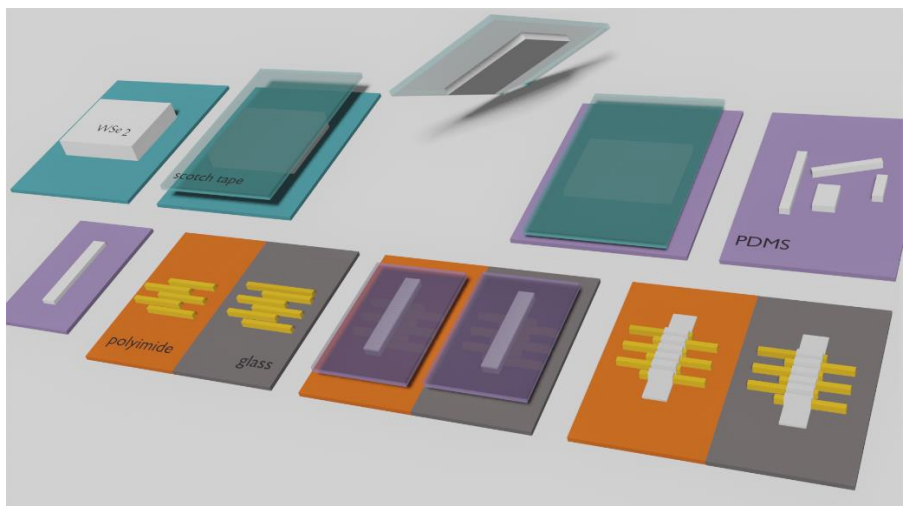

**Figure S6.** Mechanical exfoliation and stamping process. Using scotch tape (Scotch Magic™ Tape), flakes were exfoliated from the WSe$_2$ crystal and transferred to a polydimethylsiloxane (PDMS) stamp (PF Gel Film®, Teltec GmbH). Flakes were selected using an optical microscope and the PDMS stamp was trimmed with a scalpel to remove excess flakes. The PDMS stamp was brought into contact with the substrate using micromanipulator screws on the light microscope setup allowing the flake to be transferred.